\documentclass[aps, prl, 10pt, twocolumn, superscriptaddress]{revtex4-1}

\usepackage{graphicx}
\usepackage{amsmath,amssymb,amsfonts}
\usepackage[usenames, dvipsnames]{color}
\usepackage[percent]{overpic}
\usepackage{braket}

\begin{document}
\title{Extractable work, the role of correlations, and asymptotic freedom in quantum batteries}
\author{Gian Marcello Andolina}
\affiliation{NEST, Scuola Normale Superiore, I-56126 Pisa,~Italy}
\affiliation{Istituto Italiano di Tecnologia, Graphene Labs, Via Morego 30, I-16163 Genova,~Italy}
\author{Maximilian Keck}
\affiliation{NEST, Scuola Normale Superiore and Istituto Nanoscienze-CNR, I-56126 Pisa,~Italy}
\author{Andrea Mari}
\affiliation{NEST,  Scuola Normale Superiore and Istituto Nanoscienze-CNR, I-56126 Pisa,~Italy}
\author{Michele Campisi}
\affiliation{Department of Physics and Astronomy, University of Florence, Via Sansone 1, I-50019 Sesto Fiorentino (FI),~Italy}
\affiliation{INFN Sezione di Firenze, via G.Sansone 1, I-50019 Sesto Fiorentino (FI),~Italy}
\author{Vittorio Giovannetti}
\affiliation{NEST, Scuola Normale Superiore and Istituto Nanoscienze-CNR, I-56126 Pisa,~Italy}
\author{Marco Polini}
\affiliation{Istituto Italiano di Tecnologia, Graphene Labs, Via Morego 30, I-16163 Genova,~Italy}
\begin{abstract}
We investigate a quantum battery made of $N$ two-level systems, which is charged by an optical mode via an energy-conserving interaction. 
We quantify the fraction of energy stored in the battery that can be extracted in order to perform thermodynamic work. We first demonstrate that this quantity is highly reduced by the presence of correlations between the charger and the battery or between the subsystems composing the battery. We then show that the correlation-induced suppression of extractable energy, however, can be mitigated by preparing the charger in a coherent optical state. We conclude by proving that the charger-battery system is asymptotically free of such locking correlations in the $N \rightarrow \infty$ limit.
\end{abstract}


\maketitle

{\it Introduction.}---The possibility of using quantum phenomena for technological purposes is currently
a very active research field.  
In this context, an interesting research topic is that of  ``quantum batteries'' (QBs)~\cite{Alicki13, Hovhannisyan13, Binder15, Campaioli17, Ferraro17, Le17,Andolina18,Campaioli18,Serra18}, i.e.~quantum mechanical systems which behave as efficient energy storage devices. This is motivated by the fact that genuine quantum effects, such as entanglement or squeezing, can typically boost the performances of  classical protocols, e.g.~by speeding up the underlying dynamics~\cite{Defner17, Giovannetti2003a}.
The advantage provided by quantum correlations in the charging (or discharging) process of a QB has been discussed in a fully abstract fashion~\cite{Alicki13, Hovhannisyan13, Binder15, Campaioli17} and, more recently, for concrete models that could be implemented in the laboratory~\cite{Ferraro17, Le17,Andolina18}.
Up to now, research efforts have been mostly focused on maximizing the stored energy, minimizing the charging time or maximizing the average charging power~\cite{Binder15, Campaioli17, Ferraro17, Le17,Andolina18}. A ``good'' QB, however, should not only store a relevant amount of energy, but also have the capability to fully deliver such energy in a useful way which, said in thermodynamic terms, is the capability of performing work. This observation is not a negligible subtlety, since in quantum information theory it is well known that correlations and entanglement may induce limitations on the task of energy extraction~\cite{Oppenheim2002,Vitagliano18,Alicki13,Goold2016,Manabendra17,Zambrini18}. We are therefore naturally led to face a somewhat frustrating situation in which quantum correlations have simultaneously both a positive and a negative effect in the process of energy storage. On one hand, they can speed up the charging time of QBs, while, on the other hand, they can pose a severe limit on the work that can be actually extracted from it. 

\begin{figure}[t]
\centering
\begin{overpic}[width=0.9\columnwidth]{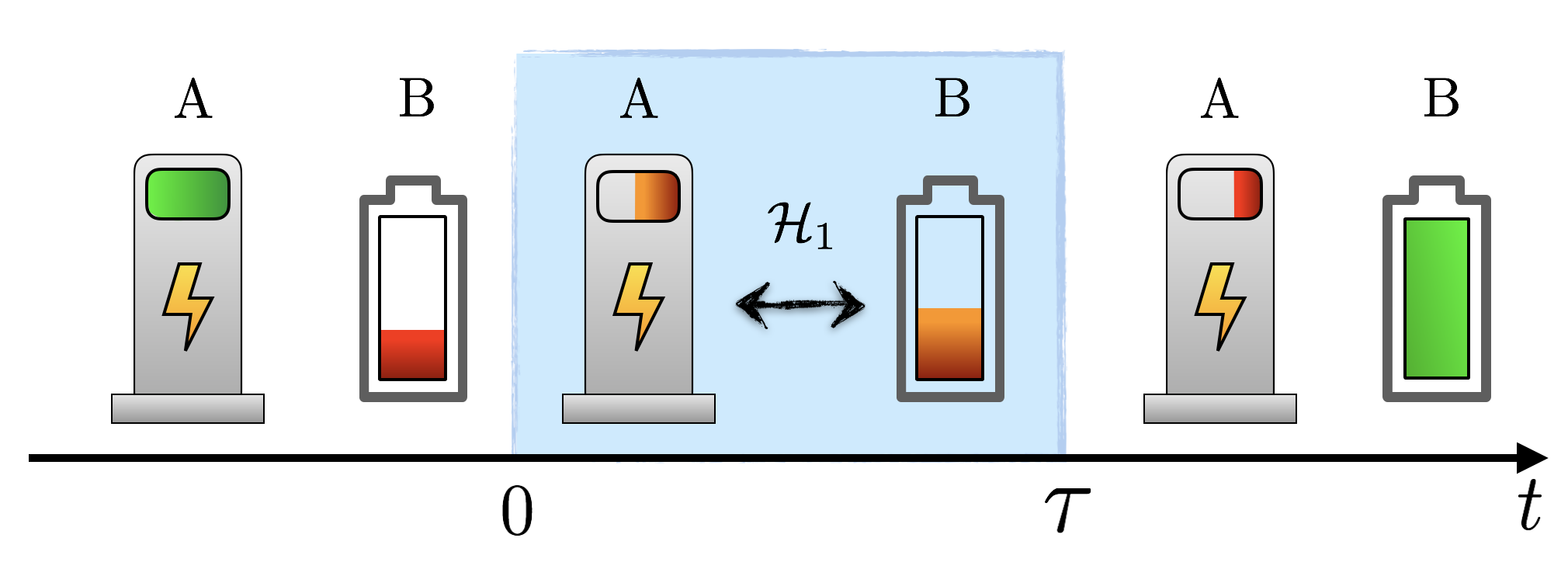}\put(4,31){\normalsize }\end{overpic}
\caption{(Color online) The charging protocol of a quantum battery. At time $t<0$ the two systems A (i.e.~the charger) and B (i.e.~the battery) do not interact and cannot exchange energy. In the time interval $0<t<\tau$, the coupling Hamiltonian $\mathcal{H}_1$ is switched on and the two subsystems interact with a coupling strength $g$. Finally, the interaction is switched off at time $\tau$ and, after that, the energy stored in the battery B, $E_{\rm B} (\tau)$, is conserved. \label{fig:System}
}
\end{figure}
In this work we shed some light on the competition between the aforementioned positive and negative aspects of quantum correlations,
by analyzing the case of $N$ two-level systems (qubits) charged via a single optical mode, the so-called Tavis-Cummings model \cite{Tavis68, Tavis69},
which is known to provide an effective description  of experimentally feasible many-body systems in circuit-QED~\cite{Fink09,Yang18,Leek09,Lolli15}.
Our findings show that in the case of QBs involving a small number of qubits the energy locked by correlations can be large and must be taken into account for a rigorous and fair analysis of the performance of the QB itself. Luckily, however, this negative effect can be strongly reduced by an optimization over the initial state of the charging system, i.e.~by properly preparing the initial state of the charger. Moreover, in the thermodynamic $N\rightarrow \infty$ limit of many qubits, the fraction of locked energy becomes negligible, independent of the initial state of the charger. We argue that this is a general property of quantum charging processes of closed Hamiltonian systems, which can be applied to other schemes (e.g.~those analyzed in Ref.~\cite{Andolina18}) beyond the specific setup presented here, being ultimately
 linked to the integrability of the dynamics and not depending on the details of the latter.

{\it Mean energy versus extractable work.}---We start by defining a general model for the charging process of a QB, schematically represented in Fig.~\ref{fig:System}.
Here a first quantum system $\rm A$ acts as the energy ``charger''  for a second quantum system $\rm B$ that instead acts as 
the battery of the model. They are characterized by local Hamiltonians 
$\mathcal{H}_{\rm A}$ and $\mathcal{H}_{\rm B}$ respectively, which, for the sake of convenience, are both selected to have zero ground-state energy.
Later on we shall also assume  ${\rm B}$ to be composed by $N$ non-mutually interacting elements: for the moment however this assumption is
not relevant, and we don't invoke it yet. At time $t=0$ the system starts in a pure factorized state  $| \psi\rangle_{\rm A} \otimes |0\rangle_{\rm B}$, with  $|0\rangle_{\rm B}$ being the ground state of $\mathcal{H}_{\rm B}$, and $| \psi\rangle_{\rm A}$ having mean local energy $E_{\rm A}(0) :={_{\rm A} \langle} \psi|\mathcal{H}_{\rm A}|\psi\rangle_{\rm A} >0$.
By switching on a coupling Hamiltonian $\mathcal{H}_1$ between the two systems, our aim is to transfer as much energy as  possible from ${\rm A}$ to ${\rm B}$, in some 
finite time interval $\tau$, the charging time of the protocol.
For this purpose, we write the global Hamiltonian of the model as
\begin{equation}
\label{eq:protocol}
\mathcal{H}(t)\equiv \mathcal{H}_{\rm A}+\mathcal{H}_{\rm B}+\lambda(t)\mathcal{H}_1~,
\end{equation}
where  $\lambda(t)$ is a classical parameter that represents the external
control we exert on the system, and which we assume to be given by
a step function  equal to $1$ for $t\in[0,\tau]$ and zero elsewhere. Accordingly, indicating with $|\psi(t) \rangle_{\rm AB}$ the evolved
state of the system at time $t$, its total energy $E(t):= {_{\rm AB}\langle} \psi(t) |\mathcal{H}(t)| 
\psi(t) \rangle_{\rm AB}$ is constant at all times with the exception of the switching points, $t=0$ and $t=\tau$, where some non-zero energy can be passed on ${\rm AB}$ by the external control. (See Ref.~\cite{Andolina18} for a detailed analysis on the energy cost of modulating the interaction.)
For the sake of simplicity, we set these contributions equal to zero by assuming $\mathcal{H}_{1}$ to commute with the local terms $\mathcal{H}_{\rm A}+\mathcal{H}_{\rm B}$~\cite{Strasberg16}. Under this condition, the  energy that shifts from  ${\rm A}$ to ${\rm B}$ can be expressed in terms  of the mean local energy of the battery at the end of the protocol, i.e.~the quantity
\begin{equation}
E_{\rm B}(\tau)\equiv {\rm tr}[\mathcal{H}_{\rm B} \rho_{\rm B}(\tau)]~,\label{stored energy}
\end{equation}
$\rho_{\rm B}(\tau)$
being the reduced density matrix of the battery at time $\tau$.
The next question to ask is which part of  $E_{\rm B}(\tau)$
can be extracted from ${\rm B}$  without having access to the charger  (a reasonable scenario in any relevant practical applications where 
the charger A is not available to the end user), and what is instead locked by the  correlations ${\rm AB}$ have established during the charging process.
A proper measure for this quantity is provided by the ergotropy~\cite{Allahverdyan04} of the state $\rho_{\rm B}(\tau)$. We remind that given a quantum system $\rm X$ characterized by a local  Hamiltonian $\mathcal H$, the ergotropy $\mathcal{E}(\rho,{\mathcal H})$ is a functional which measures the 
maximum amount of energy that can be extracted from a density matrix $\rho$ of $\rm X$ without wasting into heat. A closed expression for this quantity can be obtained in terms of the difference 
\begin{equation}\label{eq:ergotropy2}
\mathcal{E}(\rho,{\cal H})=E(\rho)-E(\tilde{\rho})\;,
\end{equation}
between the mean energy  $E(\rho) = {\rm tr}[\mathcal{H} \rho]$  of the state $\rho$ andthat, $E(\tilde{\rho}) = {\rm tr}[\mathcal{H} \tilde{\rho}]$, of the passive counterpart $\tilde{\rho}$ of $\rho$~\cite{Lenard1978,Pusz1978,Allahverdyan04,Lorch18,Scarani18,Kurizki18,Plastina17}. The latter is defined as the density matrix of $\rm X$ which is diagonal on the eigenbasis of ${\cal H}$ and whose eigenvalues correspond to a proper reordering of those of $\rho$, i.e.~$\tilde{\rho}=\sum_n r_n \ket{\epsilon_n}\bra{\epsilon_n}$ with 
$\rho=\sum_n r_n\ket{r_n}\bra{r_n} $, $\mathcal{H}=\sum_n \epsilon_n\ket{\epsilon_n}\bra{\epsilon_n}$, with  $r_0\geq r_1 \geq \cdots$ and $\epsilon_0\leq \epsilon_1 \leq \cdots$, yielding 
$E(\tilde{\rho})=\sum_{n} r_n \epsilon_n$.
Notice that, if we set the ground-state energy to zero ($\epsilon_0=0$) and if the state is pure, then $E(\tilde{\rho})=0$ and the ergotropy coincides with the mean energy of $\rho$, i.e.~$\mathcal{E}(\rho,{\cal H})=E(\rho)$.
On the contrary, if the state is mixed, the extractable work is in general smaller than the mean energy, i.e.~$\mathcal{E}(\rho,{\cal H}) < E(\rho)$. 
Since, in the problem at hand, the global system dynamics of $\rm AB$ is unitary and the initial state $\rho_{\rm AB}(0)$ of the charger-battery system is pure, $\rho_{\rm AB}(t)$ remains pure at all times. However, the local state of the battery $\rho_{\rm B}(\tau)$ will be in general mixed because of its entanglement with the charger introducing a non-trivial gap between its ergotropy 
\begin{eqnarray}
\mathcal{E}_{\rm B}(\tau)\equiv \mathcal{E}(\rho_{\rm B}(\tau),{\mathcal H}_{\rm B})~,\label{ERGO}
\end{eqnarray}
and the energy ${E}_{\rm B}(\tau)$ it stores at the end of the charging process, see Eq.~(\ref{stored energy}). As we will show below, the former can be much smaller than the latter for the experimentally relevant case of a system composed by a small number of battery elements~\cite{Stockklauser17,Samkharadze18}.

\begin{figure}[t]
\vspace{1.5em}
\centering \includegraphics[width=\columnwidth]{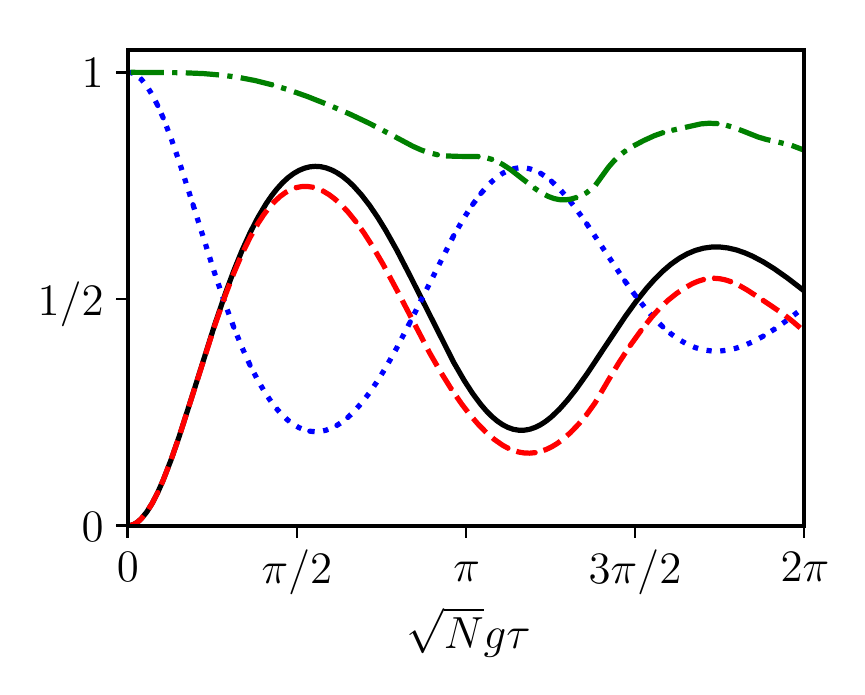}
\caption{(Color online) The energy $E^{(N)}_{\rm B} (\tau)$ (solid black line), the ergotropy $\mathcal{E}^{(N)}_{\rm B} (\tau)$ (dashed red line), the energy $E^{(N)}_{\rm A} (\tau)$ (dotted blue line), and the ratio $\mathcal{E}^{(N)}_{\rm B} (\tau)/E^{(N)}_{\rm B} (\tau)$ (dash-dotted green line) are shown as functions of $\sqrt{N}g \tau$. All quantities are measured in units of $N\omega_0$. Numerical results in this figure have been obtained by choosing a coherent state for $N=8$.\label{fig:ErgTau}}   
\end{figure}

{\it Results.}---For the sake of concreteness and the feasibility of its experimental realization, in the remaining of this work we focus on a definite model in which the charger ${\rm A}$ is a photonic cavity coupled via energy-conserving terms to a array  of $N$ non-mutually interacting qubits that act as the battery~${\rm B}$~\cite{Ferraro17}. The microscopic Hamiltonian is therefore the one of the Tavis-Cummings model~\cite{Tavis68, Tavis69}: $\mathcal{H}_{\rm A}=\omega_0 a^\dagger a$, 
$\mathcal{H}_{\rm B}=\omega_0\sum_{i=1}^{N} \sigma_{i}^{+}\sigma_{i}^{-}$, $\mathcal{H}_{1}=g \sum_{i=1}^N (a \sigma_{i}^{+} + a^\dagger\sigma_{i}^{-})$, 
where $a\,(a^\dagger)$ is a bosonic annihilation (creation) operator, $\sigma_{i}^{\pm}$ are raising/lowering spin operators for the $i$-th qubit,
$\omega_0$ is the characteristic frequency of both subsystems, and $g$ the coupling strength ($\hbar=1$ throughout this work).
In this setting we compare the final maximum extractable work measured by the ergotropy ${\mathcal E}^{(N)}_{\rm B}(\tau)$ and the mean energy $E^{(N)}_{\rm B}(\tau)$ of the battery with respect to different initial states $\ket{\psi}_{\rm A}$ of the charger (the label $N$ being added to  put emphasis on the size  of the $\rm B$ system). We restrict the analysis to three typical quantum optical states~\cite{WallsMilburn2007}: a Fock state, a coherent state, and a squeezed vacuum state, all having the same input energy $E_{\rm A}^{(N)}(0)$, which we set equal to $N \omega_0$ in order to ensure that it matches the full energy capacity of the battery.
In Fig.~\ref{fig:ErgTau} we show the stored energy $E^{(N)}_{\rm B}(\tau)$, the energy of the charger $E^{(N)}_{\rm A}(\tau)\equiv \mbox{tr}[ {\mathcal H}_{\rm A} \rho_{\rm A}(\tau)]$, and ergotropy $\mathcal{E}^{(N)}_{\rm B}(\tau)$ as functions of the duration  $\tau$ of the charging protocol, for the case of the input  coherent state. We clearly see that for $\sqrt{N}g \tau \lesssim \pi/4 $ the difference between ergotropy and energy is relatively small. Conversely, correlations that emerge between A and B at long times yield an energy and an ergotropy that are significantly different.

We now focus on the main point of this work, i.e.~a comparison between the fraction of extractable work with respect to the total mean energy of the battery. 
Consistently with previous approaches already used in the literature~\cite{Binder15,Ferraro17}, we fix the duration of the protocol to the value $\tau=\bar{\tau}$ which ensures the maximum value for the average charging power $P_{\rm B}^{(N)}(\tau) \equiv E_{\rm B}^{(N)}(\tau)/\tau$, i.e.~$P_{\rm B}^{(N)}(\tau) \le P_{\rm B}^{(N)}(\bar{\tau})$.

As explicitly discussed in Ref.~\cite{SOM}, we start by observing that {\it all} initial states exhibit
the same  $P_{\rm B}^{(N)}({\bar \tau}) \propto {N}^{3/2}$ scaling reported in Ref.~\cite{Ferraro17}, where only Fock states were considered. 
This corresponds to a $\bar{\tau}\propto 1/\sqrt{N}$ collective speed-up of the charging time, which is independent of the initial state of A, and valid, in particular, for a semi-classical coherent state. Highly non-classical initial states are therefore not necessary for optimizing the charging part of the protocol.

\begin{figure}[t]
\centering
\vspace{1.5em}
\begin{overpic}[width=\columnwidth]{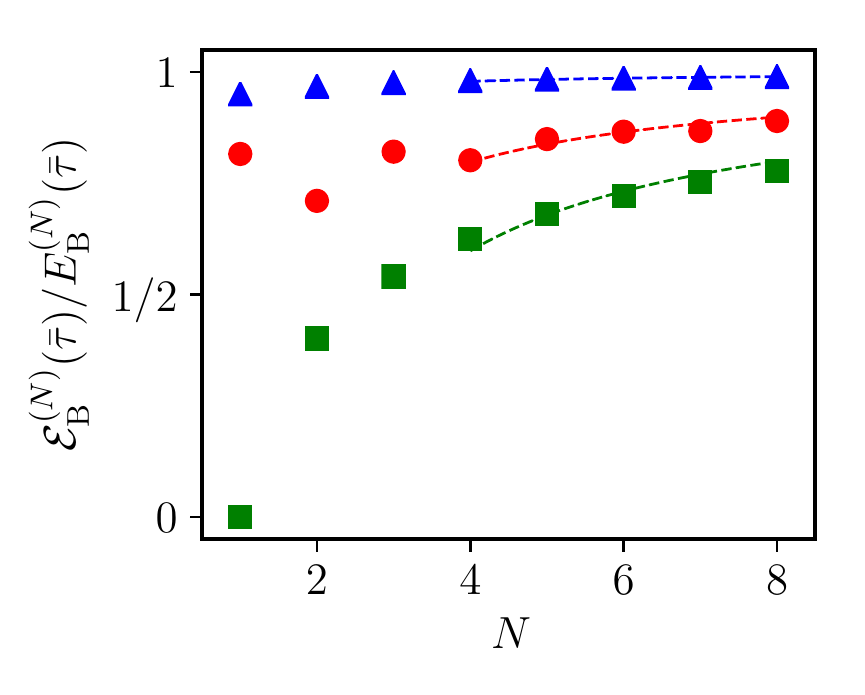}\put(4,80){\normalsize (a)}\end{overpic}
\begin{overpic}[width=\columnwidth]{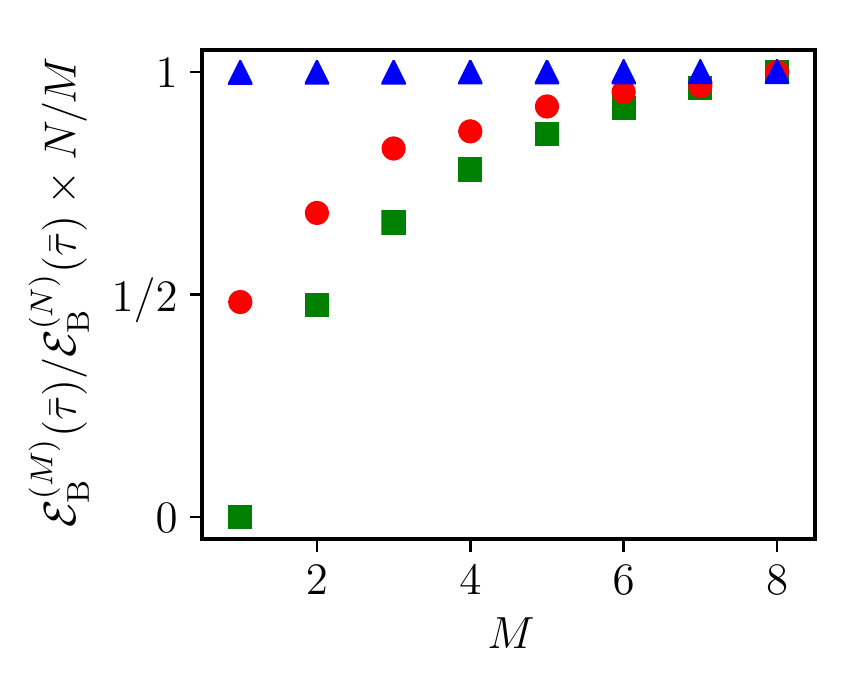}\put(4,80){\normalsize (b)}\end{overpic}
\caption{(Color online) Panel (a) The ratio $\mathcal{E}^{(N)}_{\rm B} (\bar{\tau})/E^{(N)}_{\rm B} (\bar{\tau})$ as a function of $N$ for three initial states of the charger: a Fock state (red circles), a coherent state (blue triangles), and a squeezed state (green squares). We fit the last five points of all sets of numerical data with curves that converge to $1$ with a $1/N$ scaling (dashed lines). Panel (b) The quantity $\big[\mathcal{E}^{(M)}_{\rm B} (\bar{\tau})/M \big] / \big [E^{(N)}_{\rm B} (\bar{\tau})/ N\big]$ as a function of $M\leq N$. Color coding as in panel (a). Data in panel (b) have been obtained by setting $N=8$.  \label{fig:Scaling}}
\end{figure}

Next, in Fig.~\ref{fig:Scaling}(a) we illustrate the dependence of the ratio $\mathcal{E}^{(N)}_{\rm B} (\bar{\tau})/E^{(N)}_{\rm B} (\bar{\tau})$ on the number $N$ of qubits and for the three selected initial states.
We clearly see two important facts: (i) for small $N$, the extractable work can be much smaller than the mean energy of the battery and coherent input states appear to be optimal; (ii) for large values of $N$, almost all the mean energy of the battery becomes extractable, a result which justifies a posteriori previous asymptotic approaches~\cite{Ferraro17,Binder15} to QBs in which only the mean energy was considered as a figure of merit. 
Fig.~\ref{fig:Scaling}(b) shows the amount of energy that can be extracted from a fraction of $M\leq N$ qubits (and normalized by $M$) divided by the same quantity evaluated for all $N$ qubits (and normalized by $N$), i.e.~$\big[\mathcal{E}^{(M)}_{\rm B} (\bar{\tau})/M \big] /  \big [\mathcal E^{(N)}_{\rm B} (\bar{\tau})/ N\big]$. This ratio describes the fraction of energy that can be extracted when only operations on a subset of $M$ qubits are allowed. This is of interest because performing operations on all qubits may be experimentally challenging. Our results show, however, that this is in general not necessary. Indeed, our illustrative results for $N=8$ demonstrate that operating on a subset of just $M = 4$ is already sufficient to extract~more than $\approx 3/4$ of all the available work. We further note that also in this case the coherent states are optimal. The fraction of extractable work from these initial states is weakly affected by the limitation to local operations on $M\leq N$ qubits, and is practically constant and close to $1$. These makes coherent states  ideal initial charging states for QBs ---see also Ref.~\cite{SOM}.

{\it Discussion and summary.}---  We now comment on the two main results emerging from our numerical analysis, i.e.~the optimality of coherent states for small $N$ and the asymptotic freedom of the charger-battery system from locking correlations in the $N \gg 1$ limit.

Regarding the first issue, we observe that from the ergotropy definition~(\ref{eq:ergotropy2}), it is clear that the more mixed a state is, the more difficult it is to extract its energy, a fact which is analogous to the difficulty of extracting work from a classical thermodynamic system with large entropy.  Since in our model  the joint AB  state  is pure, the entropy of the reduced density matrix of the battery is a consequence of its entanglement with the charger. We can therefore say that, for what concerns the capability of work extraction, it is convenient to produce as little entanglement as possible between the charger and the battery. From this argument, we naturally conclude that highly non-classical initial states of the charger (such as Fock or squeezed states), which induce a complex and entangling dynamics, are not optimal for work extraction. On the contrary, we expect semi-classical states like coherent states, which are well known in quantum optics for producing small entanglement under energy-conserving interactions, to be optimal for maximizing the final ergotropy of the battery (while maintaining the collective speed-up of the charging time). This argument provides a simple yet natural qualitative explanation of our numerical results. 

For what concerns instead the asymptotic freedom from locking correlations in the $N\rightarrow \infty$ limit, we
argue that this is not a peculiar feature of our model but rather a much more universal fact 
 that applies to all those systems whose dynamics is restricted to a small part of the Hilbert space, a phenomenon intrinsically connected with the integrability of the model. In order to understand this point we start again  from our previous observation that the charger-battery entanglement is the main limiting factor for the task of work extraction. It is well known that the entanglement entropy of the subsystems of an integrable system usually fails to scale with their size. This phenomenon is also known under the name of {\it area law}~\cite{eisert_rmp_2010,goold_prb_2015,nandkishore_annurev_2015}. On the contrary, the energy is an extensive quantity, which grows linearly with the size of our battery. For this reason, we expect that the relative ratio between the locked and the extractable energy is negligible in the $N \rightarrow \infty$ limit.
A way to put this observation on a more rigorous ground is via a
result we prove in Sect.~IV of Ref.~\cite{SOM}: namely that 
if the system B is composed of $N$ resonant qubits and 
the number of non-null eigenvalues of the density matrix $\rho_{\rm B}$ scales polynomially in $N$,  
than 
all its energy is accessible in the thermodynamic limit, i.e. 
\begin{equation}\label{eq:limit}
\lim_{N \rightarrow \infty} {\mathcal{E}^{(N)}_{\rm B}}/{E^{(N)}_{\rm B}} = 1~,
\end{equation}
the limit being achieved with a finite-size $1/N$  scaling, as in Fig.~\ref{fig:Scaling}(a).

Now, one can identify at least two relevant classes of models which fulfil the requirements listed above.
The first is represented by  systems which, as our integrable~\cite{bogoliubov_jphysA_1996} 
 Tavis-Cummings QB model, are characterized by  energy preserving interactions, i.e.~$[\mathcal{H}_1,\mathcal{H}_0]=0$,  and  which have  a single charger A with a not highly degenerate spectrum  and initialized into an input configuration  with a sufficiently well behaved energy distribution  (e.g.~a Fock or a  coherent state).
In this case, assuming as usual the initial mean energy of A to be proportional to $N$,  the number of relevant eigenvalues of its density matrix $\rho_{\rm A}(\tau)$ at the end of the charging process
 will be upper bounded by a quantity $d$ that
scales at most polynomially with $N$. (As a matter of fact, for the
Tavis-Cummings QB model the scaling of $d$  is indeed  linear with $N$, see e.g.~Refs.~\cite{Ferraro17,Bastarrachea11}.) This is a simple consequence of the fact that the energy of A can only be reduced by the interaction with the battery, initially in its ground state.  
Since the global state of the complete system is pure, the spectrum of 
$\rho_{\rm A}(\tau)$  will be equal to the spectrum of $\rho_{\rm B}(\tau)$~\cite{Nielsen_and_Chuang} 
making the number of  its non-negligible eigenvalues also equal to $d$, and hence ultimately leading to Eq.~(\ref{eq:limit}).

The second class of models for which we expect Eq.~(\ref{eq:limit}) to hold, are those
where the dynamics of the QB is restricted to a  small subspace 
of the entire exponentially large Hilbert space due to the conservation of some operator and the form of the initial state. A notable example is the Dicke model~\cite{Dicke54}, which exhibits conservation of  $(J^{(N)})^2$.  In this case, the initial state for the battery has a definite eigenvalue for $(J^{(N)})^2$, namely $J=N/2$, and hence all the dynamics of B lies in the subspace with a definite $J$ leading once more to Eq.~(\ref{eq:limit}), as we explicitly 
show, via numerical analysis in Sect.~V of Ref.~\cite{SOM}.

In summary, by studying a physically well motived  QB model, 
we found that, for a small number of batteries (as in current state-of-the-art solid-state technology~\cite{Fink09,Yang18,Leek09,Lolli15}),  the extractable energy can be significantly smaller than the  mean energy stored in the devices. This negative effect strongly depends on the choice of the initial state of the charger and we found that coherent states are optimal for mitigating this phenomenon. For a large number of the batteries, instead, we found that 
  the  extractable energy  converges to the stored energy. 
  We also argued that this a rather universal phenomenon characterizing all charger-battery systems in which the amount of entanglement is not extensive with respect to the size $N$ of the battery.

{\it Acknowledgments.}---Numerical work has been performed by using the Python toolbox
QuTiP2~\cite{QuTip}. We wish to thank D. Farina, D. Ferraro, P.A. Erdman, V. Cavina, and F.M.D.  Pellegrino for useful discussions.

\clearpage 
\setcounter{section}{0}
\setcounter{equation}{0}%
\setcounter{figure}{0}%
\setcounter{table}{0}%

\setcounter{page}{1}

\renewcommand{\thetable}{S\arabic{table}}
\renewcommand{\theequation}{S\arabic{equation}}
\renewcommand{\thefigure}{S\arabic{figure}}
\renewcommand{\bibnumfmt}[1]{[S#1]}
\renewcommand{\citenumfont}[1]{S#1}

\onecolumngrid

\begin{center}
\textbf{\Large Supplemental Material for ``Extractable work, the role of correlations, and asymptotic freedom in quantum batteries''}
\bigskip

Gian Marcello Andolina,$^{1,\,2}$
Maximilian Keck,$^3$
Andrea Mari,$^3$
Michele Campisi,$^{4,\,5}$
Vittorio Giovannetti,$^3$ and
Marco Polini$^1$

\bigskip

$^1$\!{\it Istituto Italiano di Tecnologia, Graphene Labs, Via Morego 30, I-16163 Genova,~Italy}

$^2$\!{\it NEST, Scuola Normale Superiore, I-56126 Pisa,~Italy}

$^3$\!{\it NEST, Scuola Normale Superiore and Istituto Nanoscienze-CNR, I-56126 Pisa,~Italy}

$^4$\!{\it Department of Physics and Astronomy, University of Florence, Via Sansone 1, I-50019 Sesto Fiorentino (FI),~Italy}

$^5$\!{\it INFN Sezione di Firenze, via G.Sansone 1, I-50019 Sesto Fiorentino (FI),~Italy}

\bigskip

In this Supplemental Material we provide additional information on the explicit form of the initial states mentioned in the main text and the scaling of the average charging power with $N$ for these states. Finally, we also elaborate on why the coherent state is optimal for the ergotropy.
\end{center}

\maketitle

\appendix
\twocolumngrid

\section{Explicit form of the three initial states of the charger}
\label{Appendix:InitialStates}

We here provide the explicit form of the three initial states studied in our work, i.e.~a Fock state, a coherent state, and a squeezed state:
\begin{eqnarray}\label{initialStates}
\ket{n}_{\rm A}&=&\frac{(a^\dagger)^n}{\sqrt{n!}}\ket{0}~, \nonumber\\ 
\ket{\alpha}_{\rm A}&=&\exp\Big(\alpha a^\dagger-\alpha^* a\Big)\ket{0}~, \nonumber\\
\ket{z}_{\rm A}&=&\exp\Big(\frac{z a^2- z^* (a^\dagger)^2}{2} \Big)\ket{0}~,
\end{eqnarray}
where $\ket{0}$ is the vacuum of the cavity. The three parameters $n$, $\alpha$, and $z$, are fixed by the requirement to have  input energy  equal to $E_{\rm A}^{(N)}(0)=N\omega_0$. We therefore have $n=N$, $\alpha=\sqrt{N}$, and $z={\rm arcsinh}(\sqrt{N})$.

\section{Scaling of the maximum average charging power}
\label{Appendix:Power}

\begin{figure}[t]
\vspace{4mm}
\centering
\begin{overpic}[width=1\columnwidth]{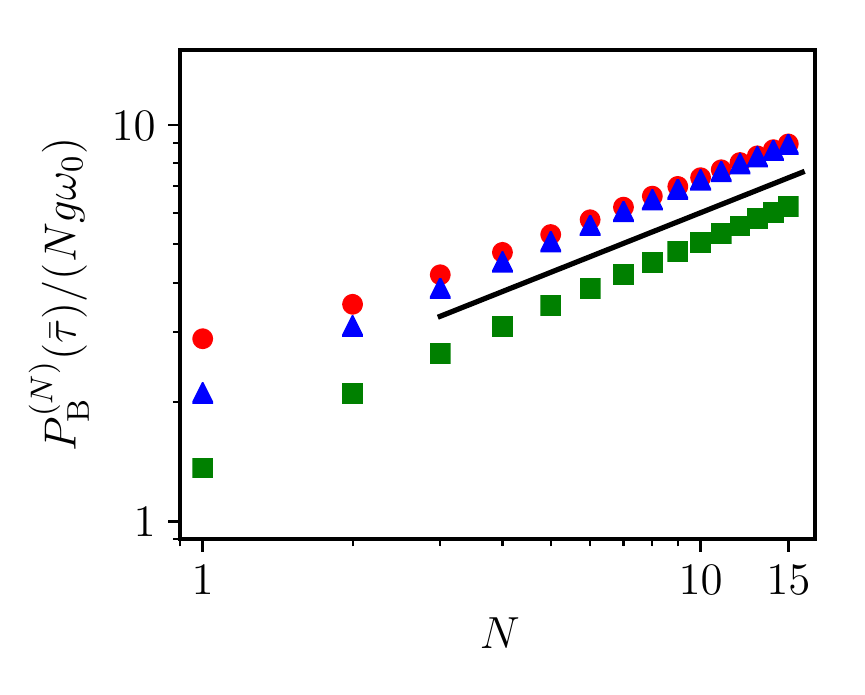}\put(4,80){\normalsize (a)}\end{overpic}
\begin{overpic}[width=1\columnwidth]{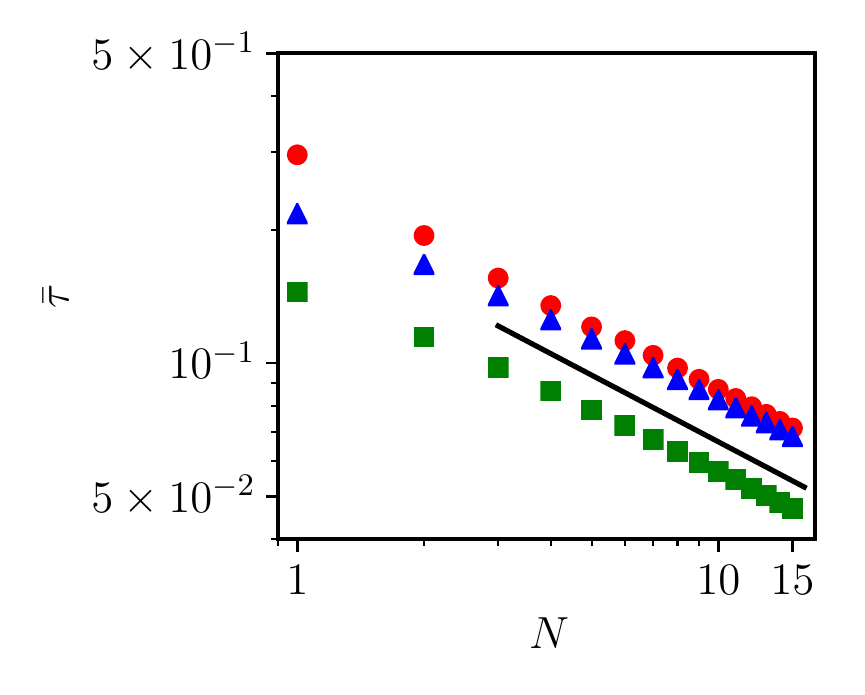}\put(4,80){\normalsize (b)}\end{overpic}
\caption{(Color online)Panel (a) The maximum average charging power $P^{(N)}_{\rm B}(\bar{\tau})$ (in units of $N g \omega_{0}$) is plotted as a function of $N$ in a log-log scale. Different symbols refer to three initial states of the charger: a Fock state (red circles), a coherent state (blue triangles), and a squeezed state (green squares).We plot a black solid line with slope $1/2$ in the log-log scale, indicating the $\sqrt{N}$ scaling.Panel (b) The time correspondig to the maximum average charging power average charging power  $\bar{\tau}$  is plotted as a function of $N$ in a log-log scale. We plot a black solid line with slope $-1/2$ in the log-log scale, indicating the $1/\sqrt{N}$ scaling. \label{fig:power}}
\end{figure}
Here, we study the maximum average charging power $ P^{(N)}_{\rm B}(\bar{\tau})$ as a function of $N$, for the three initial states introduced in the main text and in the previous section of this file. In Fig.~\ref{fig:power} panel (a) we report a log-log scale plot of $P^{(N)}_{\rm B}(\bar{\tau})/N$ as a function of $N$. A simple inspection of this plot shows that $P^{(N)}_{\rm B}(\bar{\tau})/N \propto \sqrt{N}$, independently of the initial state. Now, by definition, $P^{(N)}_{\rm B}(\bar{\tau})=E_{\rm B}^{(N)}(\bar{\tau})/\bar{\tau}$. Since $E_{\rm B}^{(N)}(\bar{\tau})$ is an extensive quantity, the collective advantage $P^{(N)}_{\rm B}(\bar{\tau})/N \propto \sqrt{N}$ stems from the scaling $\bar{\tau}\propto 1/\sqrt{N}$ of the optimal time. In Fig.~(\ref{fig:power}) panel (b) we report a log-log scale plot of $\bar{\tau}$ as a function of $N$, which confirm the scaling $1/\sqrt{N}$.

\section{Optimality of coherent states for work extraction} 
\label{Appendix:proof3}

\begin{figure}[t]
\vspace{2mm}
\centering
\begin{overpic}[width=1\columnwidth]{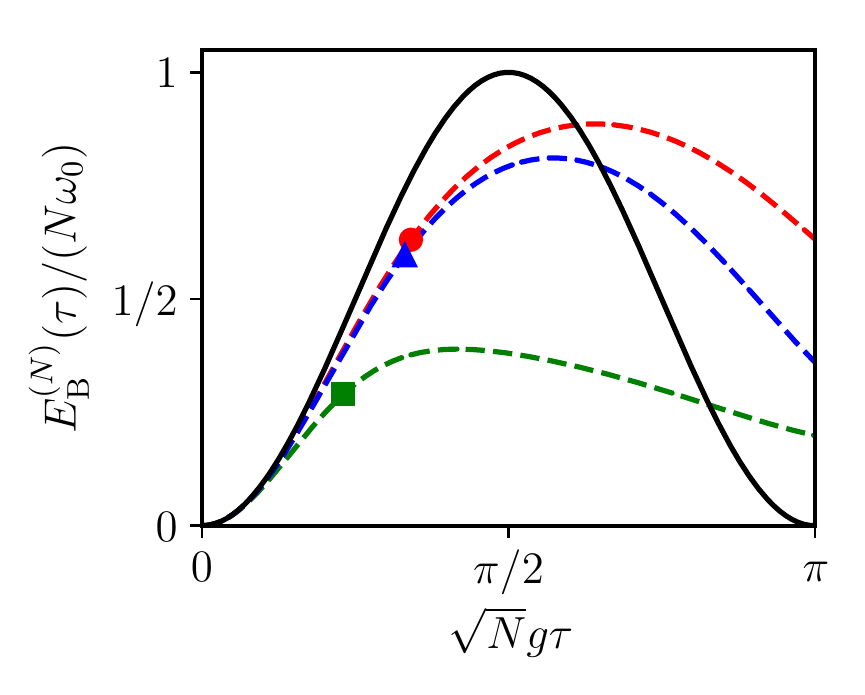}\put(4,80){\normalsize }\end{overpic}
\caption{(Color online) The stored energy $E^{(N)}_{\rm B}(\tau)$ (in units of $N\omega_{0}$) as a function of $\sqrt{N} g \tau$. The black solid line denotes the energy calculated from the approximate Hamiltonian (\ref{eqHOHO}). Dashed lines denote the energy as calculated numerically from the exact dynamics, for three initial states of the charger: a Fock state (red), a coherent state (blue), and a squeezed state (green). Filled symbols denote the value of the energy at the optimal time ${\bar \tau}$. \label{fig:proof}}
\end{figure}
In this Section we offer an argument that explains why coherent states with an high number $N$ of average excitations create weak correlations between A and B for relatively small times $\sqrt{N}g\tau\lesssim\pi/3$, B being initially in the ground state. We start by rewriting the Tavis-Cummings Hamiltonian in terms of collective spins operators:
\begin{equation}
J^{(N)}_z \equiv \frac{1}{2}\sum_{i=1}^N \sigma_{i}^{z}
\end{equation}
and
\begin{equation}
J^{(N)}_{+} \equiv \big[J^{(N)}_{-}\big]^\dagger=\sum_{i=1}^N \sigma_{i}^{+}~. 
\end{equation} 
The Hamiltonian then reads
\begin{eqnarray}
\mathcal{H}_{\rm A}&=&\omega_0 a^\dagger a~,\nonumber \\ 
\mathcal{H}_{\rm B}&=&\omega_0 \left[ J^{(N)}_z + \frac{N}{2} \right]~,\nonumber\\ 
\mathcal{H}_{1}&=&g \left[a J_{+}^{(N)}+a ^\dagger J_{-}^{(N)} \right]~.
\end{eqnarray}
We now use the Holdstein-Primakoff transformation~\cite{Holdstein40S} to express the collective spin operators in terms of auxiliary harmonic oscillator operators $b$ and $b^\dagger$: $J^{(N)}_z=(b^\dagger b-N/2)$ and  $J^{(N)}_{+}=b^\dagger\sqrt{N} \sqrt{1- b^\dagger b/N}$. If we are interested only in the first few excitations of the spectrum we can neglect terms like $b^\dagger b/N$ (since $N\gg1$). In this case, we have $J^{(N)}_{+}\approx b^\dagger \sqrt{N}$, obtaining
\begin{eqnarray}
\label{eqHOHO}
\mathcal{H}_{\rm A}&=&\omega_0 a^\dagger a~,\nonumber\\ 
\mathcal{H}_{\rm B}&=&\omega_0 b^\dagger b~, \nonumber\\
\mathcal{H}_{1}&\approx&g\sqrt{N} (a b^\dagger+a ^\dagger b)~.
\end{eqnarray}
The total Hamiltonian is now approximately the one of two harmonic oscillators coupled via a quadratic term. 
When this approximation holds, an initial coherent state remains a coherent state under time evolution, i.e.~
\begin{eqnarray}\label{psit}
\ket{\Psi(t)}&=&\exp(-i\mathcal{H}t)\ket{\sqrt{N}}_{\rm A} \otimes \ket{0}_{\rm B}\nonumber\\
&=&\ket{\sqrt{N}\cos(g_{N}t)}_{\rm A} \otimes\ket{-i\sqrt{N}\sin(g_{N}t}_{\rm B}~,
\end{eqnarray}
where $g_N=\sqrt{N}g$ and $\ket{\alpha(t)}$ is the coherent state defined by the displacement parameter $\alpha(t)$. The energy stored in B, 
as calculated from Eq.~(\ref{eqHOHO}), is $E_{\rm B}(\tau)\approx N\omega_0\sin^2(g_Nt)$ and is independent of the initial state.

The large-$N$ bosonic approximation is good only for small times, i.e.~for $\sqrt{N}g\tau\ll 1$, when the battery is poorly charged and highly excited states are empty. Furthermore, we can verify {\it a posteriori} the condition $b^\dagger b/N \ll1$ by calculating the occupation number in B within the approximation. This yields $\braket{b^\dagger b}/N=\sin^2(g_Nt)\ll 1$, which works for $\sqrt{N}g\tau\ll 1$. 

In Fig.~\ref{fig:proof} we compare the energy $E^{(N)}_{\rm B}$ calculated within the large-$N$ bosonic approximation (black solid line) with that calculated from the exact dynamics. In addition, we indicate by filled symbols the value $E^{(N)}_{\rm B}(\bar{\tau})$ evaluated at the optimal time $\bar{\tau}$. We clearly see that at the optimal time $\bar{\tau}$ the large-$N$ bosonic approximation is qualitatively correct.

\begin{figure}[t]
\centering
\vspace{1.5em}
\begin{overpic}[width=1\columnwidth]{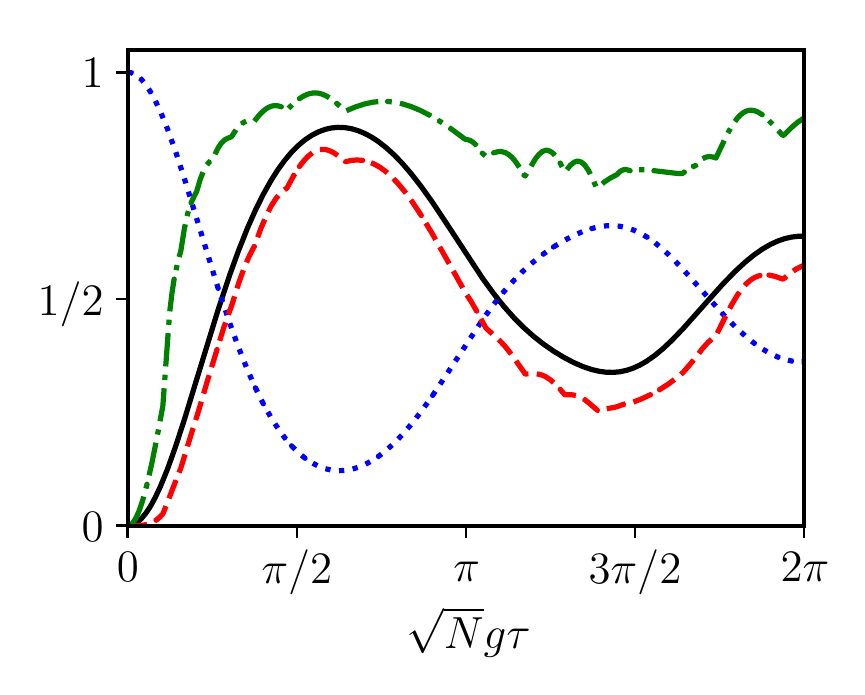}\put(4,80){\normalsize (a)}\end{overpic}
\begin{overpic}[width=1\columnwidth]{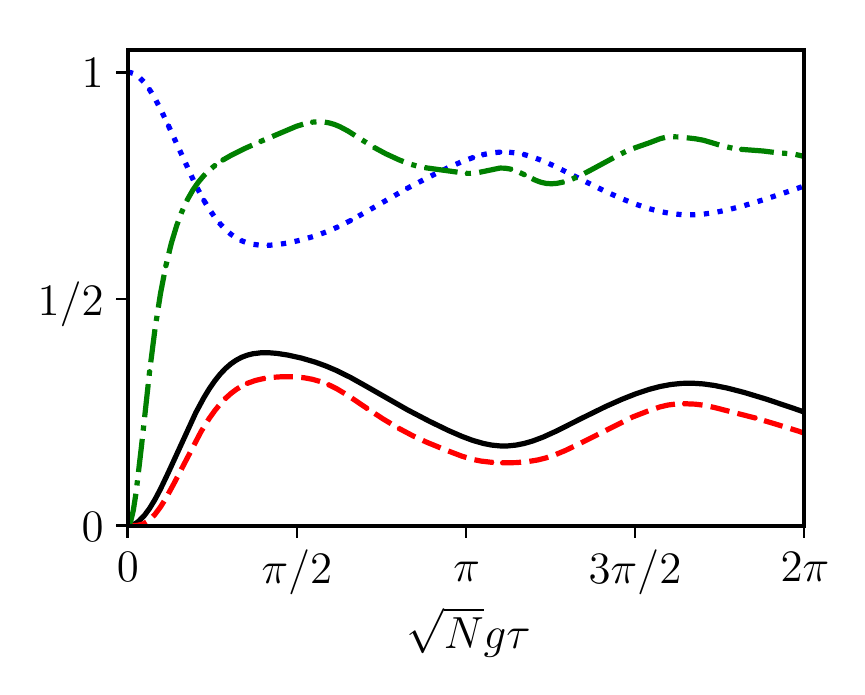}\put(4,80){\normalsize (b)}\end{overpic}
\caption{(Color online) Panel (a) The energy $E^{(N)}_{\rm B} (\tau)$ (solid black line), the ergotropy $\mathcal{E}^{(N)}_{\rm B} (\tau)$ (dashed red line), the energy $E^{(N)}_{\rm A} (\tau)$ (dotted blue line),  and the ratio $\mathcal{E}^{(N)}_{\rm B} (\tau)/E^{(N)}_{\rm B} (\tau)$(dash dotted green line)  are shown as functions of $\sqrt{N}g \tau$. All quantities are measured in units of $N\omega_0$. Numerical results in this figure have been obtained by choosing a Fock state for $N=8$. Panel (b) The same quantities for a squeezed initial state . \label{fig:Fock-Sque} }
\end{figure}

In Fig.~\ref{fig:Fock-Sque} we finally present
the energy $E^{(N)}_{\rm B} (\tau)$, the ergotropy $\mathcal{E}^{(N)}_{\rm B} (\tau)$, and the energy $E^{(N)}_{\rm A} (\tau)$   for 
 Fock (panel a) and squeezed input states (panel b). Comparing these   plots with  those obtained for the case of the coherent input  presented in  Fig.~2 of the main text,
 the optimality of this last in terms of the relation between ergotropy of B and its mean energy clearly emerges. 
   Figure.~\ref{fig:Fock-Sque} panel (b) also show that  the squeezed input exhibits a poor performance also in terms of net
   mean energy transfer. 
\section{Proof of Eq.~(6)}
\label{Appendix:proof2}

In this Section we provide a proof for Eq.~(6) of the main text, i.e.~we show that  
$\lim_{N\rightarrow \infty} \mathcal{E}^{(N)}_{\rm B}  / E^{(N)}_{\rm B} \to 1$, or more precisely, that
\begin{eqnarray}
  \label{eq:Ergto1.a}
  \lim_{N\rightarrow \infty} \frac{{E}^{(N)}_{\rm B}-\mathcal{E}^{(N)}_{\rm B}}{E^{(N)}_{\rm B}}\to 0
\end{eqnarray}
whenever  $\rho_{\rm B}({\tau})$ has a number ${\cal N}$ of non-zero eigenvalues $\lambda_i$ which is at most polynomial in $N$, i.e.~${\cal N} \leq \alpha N^k$ for some $\alpha, k > 0$. This assumption results in a von Neumann entropy $S(\rho_{\rm B}({\tau}))$ of the reduced system, which scales at most as $\log N$.

We first notice that the numerator of Eq.~(\ref{eq:Ergto1.a}) can be rewritten as ${E}^{(N)}_{\rm B}-\mathcal{E}^{(N)}_{\rm B}={\rm tr}_{\rm B}[\mathcal{H}_{\rm B}\tilde{\rho}_{\rm B}]$, where $\tilde{\rho}_{\rm B}$ is the passive state corresponding to
$\rho_{\rm B}$ and $\mathcal{H}_{\rm B}$.

We then define $\bar{\rho}_{\rm B}$ as the density matrix diagonal in the energy basis of $\mathcal{H}_{\rm B}$ 
 such that its first $\alpha N^k$ eigenvalues are non-zero and all equal, i.e.~$\bar{\lambda}_i=1/(\alpha N^k)$ for $i=1 \dots \alpha N^k$.
It is now useful to revise the concept of  majorization~\cite{Nielsen_and_ChuangS}:
a state $\rho$ majorizes another state $\rho^\prime$ (and we write $\rho\succ \rho^\prime$) if the eigenvalues of the density matrices satisfy: $\sum_{i=1}^n \lambda_i\geq\sum_{i=1}^n \lambda_i^\prime$, for all $n$, where $ \lambda_i\, (\lambda^\prime_i$) are the eigenvalues of $\rho\, (\rho^\prime$) in descending order.
Notice then that $\bar{\rho}_{\rm B}$ is a passive state and that it is majorized by $\tilde{\rho}_{\rm B}$: i.e. $\tilde{\rho}_{\rm B} \succ \bar{\rho}_{\rm B}$.

It is now useful to write
\begin{eqnarray} {\rm tr}_{\rm B}[\mathcal{H}_{\rm B}\tilde{\rho}_{\rm B}]- {\rm tr}_{\rm B}[\mathcal{H}_{\rm B}\bar{\rho}_{\rm B}] &=& \sum_i  (\epsilon^{\rm B}_{i+1}-\epsilon^{\rm B}_i) \nonumber\\
&\times&\sum_{j=1}^i(\lambda_j -\bar{\lambda}_j)~,
 \end{eqnarray}
which appears to be a negative quantity due to the fact that  the energies are ordered, i.e.~$(\epsilon^{\rm B}_{i+1}-\epsilon^{\rm B}_i)\geq 0$, and  $\tilde{\rho}_{\rm B}\succ \bar{\rho}_{\rm B}$. Accordingly, we arrive at the following inequality 
\begin{eqnarray}
{E}^{(N)}_{\rm B}-\mathcal{E}^{(N)}_{\rm B} =
 {\rm tr}_{\rm B}[\mathcal{H}_{\rm B}\tilde{\rho}_{\rm B}] \leq {\rm tr}_{\rm B}[\mathcal{H}_{\rm B}\bar{\rho}_{\rm B}]~.\label{QUA}
\end{eqnarray} 
(See also Ref.~\cite{Binder15TS}.) We now observe that the spectrum of B is highly degenerate. Since we have $N$ identical qubits, we have $N+1$  degenerate energy levels, and the $j$-th level is $\binom{N}{j}$ times degenerate. For large $N$ we can use the Stirling approximation $\binom{N}{j}\sim N^j$. If we select the level $j=k+1$ for large enough $N$ we can construct a state $\rho^{(k)}_{\rm B}$ within this degenerate subspace that populates equally $\alpha N^k$ of these $\binom{N}{k}$ states, where each state has energy $k\omega_0$. As this state has the same spectrum of $\bar{\rho}_{\rm B}$, there is a unitary operation $U_k$ such that ${\rho}^{(k)}_{\rm B}=U_k\bar{\rho}_{\rm B}U^\dagger_k$.

Now, recalling that $\bar{\rho}_{\rm B}$ is a passive state, the following inequality holds true: ${\rm tr}_{\rm B}[\mathcal{H}_{\rm B}\bar{\rho}_{\rm B}]\leq {\rm tr}_{\rm B}[\mathcal{H}_{\rm B}\rho^{(k)}_{\rm B}] = k\,\omega_0$, which inserted into Eq.~(\ref{QUA}) yields
\begin{eqnarray}
  \label{eq:Ergto1.b}
  {E}^{(N)}_{\rm B} - \mathcal{E}^{(N)}_{\rm B} \leq k \omega_0 ~.
\end{eqnarray}

Equation~(\ref{eq:Ergto1.a}) finally follows from the observation that the energy is extensive, i.e.~that $E^{(N)}_{\rm B}\sim N \omega_0$, so that
$({{E}^{(N)}_{\rm B}-\mathcal{E}^{(N)}_{\rm B}})/{E^{(N)}_{\rm B}}\sim k/N$, which goes to zero as $N\to\infty$. We also note that from this proof we can infer a convergence to unity scaling as $1/N$.

\section{Validity of our main results for the case of the Dicke model} 
\label{Appendix:Dicke}
\begin{figure}[t]
\centering
\vspace{1.5em}
\begin{overpic}[width=1\columnwidth]{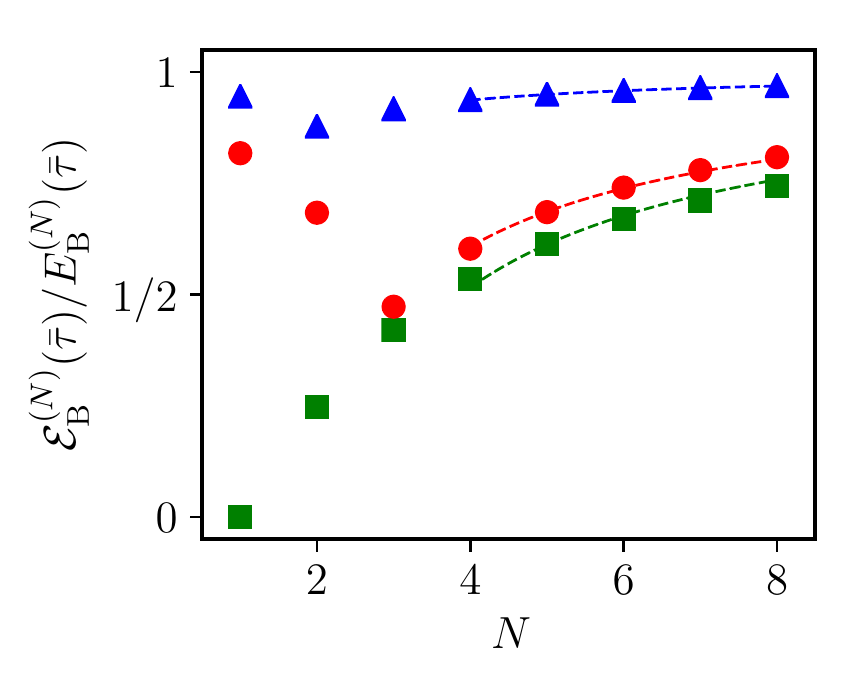}\put(4,80){\normalsize (a)}\end{overpic}
\begin{overpic}[width=1\columnwidth]{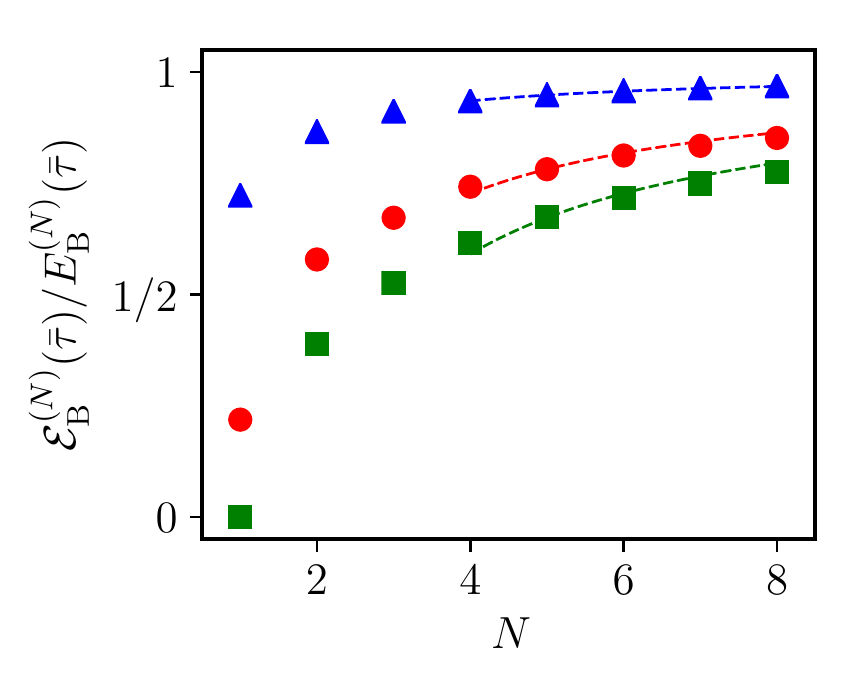}\put(4,80){\normalsize (b)}\end{overpic}
\caption{(Color online) Panel (a) The ratio $\mathcal{E}^{(N)}_{\rm B} (\bar{\tau})/E^{(N)}_{\rm B} (\bar{\tau})$ as a function of $N$ for three initial states of the charger: a Fock state (red circles), a coherent state (blue triangles), and a squeezed state (green squares). Numerical results in this figure have been obtained for the Dicke model and $g/\omega_0=0.2$. We fit the last five points of all data sets with a $1/N$ convergence to~$1$. Panel (b) Same as in panel (a) but for $g/\omega_0=2$ (the ultra-strong-coupling regime).\label{fig:ErgDicke} }
\end{figure}

In this Section we study how our analysis can be gerenalized to the Dicke model~\cite{Dicke54S,Ferraro17S}, described by the Hamiltonian
\begin{eqnarray}
  \label{HamiltonianDicke}
  \mathcal{H}_{\rm A}&=&\omega_0 ~a^\dagger a~,\\
  \mathcal{H}_{\rm B}&=&\omega_0\sum_{i=1}^{N} \sigma_{i}^{+}\sigma_{i}^{-}~,\\ 
  \mathcal{H}_{1}&=&g \sum_{i=1}^N \big(a\,+ a^\dagger\big) \sigma_{i}^{x}~,
    \label{HamiltonianDickeInteraction}
\end{eqnarray}
where all relevant operators and parameters have the save meaning as for the Tavis-Cummings model in the main text. We notice that the Tavis-Cummings Hamiltonian used in the main text can be obtained from the full Dicke Hamiltonian ~(\ref{HamiltonianDickeInteraction}) in the weak-coupling regime $g/\omega_0\ll1$, neglecting the counter-rotating terms  $ a\, \sigma_{i}^{-}$ and $ a^\dagger\, \sigma_{i}^{+}$.

Since $[ \mathcal{H}_{1}, \mathcal{H}_{0}]\neq0$ during the protocol some energy is injected via the modulation of the coupling $\lambda(t)$ and this process does not fall in the class of pure-energy-transfer protocols. (See Ref.~\cite{Andolina18S} for a detailed discussion.) While we are not explicitly studying this issue here, we show that the main results of the main text apply also to the Dicke model. We consider the same three initial optical states considered in the main text. In Fig.~\ref{fig:ErgDicke} we can clearly see that:
\begin{itemize}
\item[1)]  Coherent states are still optimal in minimizing the amount of correlations between the charger A and the battery B and hence maximize the ratio $\mathcal{E}^{(N)}_{\rm B} (\tau)/{E}^{(N)}_{\rm B} (\tau)$. We note that the amount of energy locked in correlations can be bigger than in the Tavis-Cummings case, due to the presence of counter-rotating terms.

\item[2)]  As explained in the main text, due to the conservation of the total angular momentum $J^2$, we still have the asymptotic freedoms from correlations, namely $\mathcal{E}^{(N)}_{\rm B} /{E}^{(N)}_{\rm B} \to 1$ for $N\to \infty$ also for the Dicke model.
\end{itemize}


\begin{thebibliography}{77}
%
\bibitem{Alicki13} 
R. Alicki and M. Fannes,
 \href{http://dx.doi.org/10.1103/PhysRevE.87.042123}{Phys. Rev. E~{\bf 87}, 042123 (2013)}.
%
\bibitem{Hovhannisyan13} 
K.V. Hovhannisyan, M. Perarnau-Llobet, M. Huber, and A. Ac\'in, 
\href{https://doi.org/10.1103/PhysRevLett.111.240401}{Phys. Rev. Lett.~{\bf 111}, 240201 (2013)}.
%
\bibitem{Binder15} 
F.C. Binder, S. Vinjanampathy, K. Modi, and J. Goold, 
\href{http://dx.doi.org/10.1088/1367-2630/17/7/075015}{New J. Phys.~{\bf 17}, 075015 (2015)}.
%
\bibitem{Campaioli17} 
F. Campaioli, F.A. Pollock, F.C. Binder, L. C\'{e}leri, J. Goold, S. Vinjanampathy, and K. Modi, 
\href{http://dx.doi.org/10.1103/PhysRevLett.118.150601}{Phys. Rev. Lett.~{\bf 118}, 150601 (2017)}.
%
\bibitem{Le17} 
T.P. Le, J. Levinsen, K. Modi, M. Parish, and F.A. Pollock, 
\href{https://dx.doi.org/10.1103/PhysRevA.97.022106}{Phys. Rev. A~{\bf 97}, 022106 (2018)}.
%
\bibitem{Ferraro17} 
D. Ferraro, M. Campisi, G.M. Andolina, V. Pellegrini, and 
M. Polini, \href{https://dx.doi.org/110.1103/PhysRevLett.120.117702} {Phys. Rev. Lett.~{\bf 120}, 117702 (2018)}.
%
\bibitem{Andolina18} 
G.M. Andolina, D. Farina, A. Mari, V. Pellegrini, V. Giovannetti, and M. Polini,\href{https://doi.org/10.1103/PhysRevB.98.205423} {Phys. Rev. B~{\bf 98 }, 205423 (2018)}.
%
\bibitem{Campaioli18}
For a recent review see e.g.~F. Campaioli, F.A. Pollock, and S. Vinjanampathy, \href{https://arxiv.org/abs/1805.05507}{arXiv:1805.05507}.
%
\bibitem{Serra18}
I. Henao and R.M. Serra,
\href{https://doi.org/10.1103/PhysRevE.97.062105}{Phys. Rev. E~{\bf 97}, 062105, (2018)}.
%
\bibitem{Defner17}
S. Deffner and S. Campbell, 
\href{http://dx.doi.org/10.1088/1751-8121/aa86c6}{J. Phys. A: Math. Theor.~{\bf 50}, 453001 (2017)}.
%
\bibitem{Giovannetti2003a}
 V. Giovannetti, S. Lloyd, and L. Maccone, 
 \href{https://doi.org/10.1209/epl/i2003-00418-8}{Europhys. Lett.~{ \bf 62}, 615 (2003)}.
 %
\bibitem{Oppenheim2002}
J. Oppenheim, M. Horodecki, P. Horodecki, and R. Horodecki, \href{https://doi.org/10.1103/PhysRevLett.89.180402}{Phys. Rev. Lett.~{\bf 89}, 180402 (2002)}.
%
\bibitem{Vitagliano18}
G. Vitagliano, C. Kl\"ockl, M. Huber, and N. Friis, \href{https://arxiv.org/abs/1803.06884}{arXiv:1803.06884}.
%
\bibitem{Goold2016}
J. Goold, M. Huber, A. Riera, L. del Rio, and P. Skrzypczyk, \href{https://doi.org/10.1088/1751-8113/49/14/143001}{J. Phys. A: Math. Theor.~{\bf 49}, 143001 (2016)}.
%
\bibitem{Manabendra17}
M.N. Bera, A. Riera, M. Lewenstein and A. Winter,
\href{https://doi.org/10.1038/s41467-017-02370-x}{Nat. Comm.~{\bf 8}, 2180 (2017)}.
%
\bibitem{Zambrini18}
G. Manzano, F. Plastina, and R. Zambrini,
\href{https://arxiv.org/abs/1805.08184}{arXiv:1805.08184}
%
%
\bibitem{Tavis68} 
M. Tavis and F.W. Cummings, \href{https://doi.org/10.1103/PhysRev.170.379}{Phys. Rev.~{\bf 170}, 379 (1968)}.
%
\bibitem{Tavis69} 
M. Tavis and F.W. Cummings, \href{https://doi.org/10.1103/PhysRev.188.692}{Phys. Rev.~{\bf 188}, 692 (1969)}.
%
%
\bibitem{Fink09} 
J.M. Fink, R. Bianchetti, M. Baur, M. G\"{o}ppl, L. Steffen, S. Filipp, P.J. Leek, A. Blais, and A. Wallraff, \href{http://dx.doi.org/10.1103/PhysRevLett.103.083601}{Phys. Rev. Lett.~{\bf 103}, 083601 (2009)}.
%
\bibitem{Yang18}
P. Yang, J.D. Brehm, J. Lepp\"{a}kangas, L. Guo, M. Marthaler, I. Boventer, A. Stehli, T. Wolz, A.V. Ustinov, and M. Weides,
\href{https://arxiv.org/abs/1810.00652}{arXiv:1810.00652}.
%
\bibitem{Leek09}
P.J. Leek, S. Filipp, P. Maurer, M. Baur, R. Bianchetti, J.M. Fink, M. G\"{o}ppl, L. Steffen, and A. Wallraff, 
\href{http://dx.doi.org/10.1103/PhysRevB.79.180511}{Phys. Rev. B~{\bf 79}, 180511(R) (2009)}.
%
\bibitem{Lolli15}
J. Lolli, A. Baksic, D. Nagy, Vladimir E. Manucharyan, and C. Ciuti,
\href{http://dx.doi.org/0.1103/PhysRevLett.114.183601}{Phys. Rev. Lett.~{\bf 114}, 183601 (2015)}.
%
\bibitem{Strasberg16}
P. Strasberg, G. Schaller, T. Brandes, and M. Esposito, \href{https://doi.org/10.1103/PhysRevX.7.021003}{Phys. Rev. X~{ \bf 7}, 021003 (2017)}.	
%
\bibitem{Allahverdyan04} 
A.E. Allahverdyan, R. Balian, and T.M. Nieuwenhuizen, \href{https://doi.org/10.1209/epl/i2004-10101-2}{Europhys. Lett.~{\bf 67}, 565 (2004)}.
%
\bibitem{Lenard1978} 
A. Lenard, \href{https://doi.org/10.1007/BF01011769}{J. Stat. Phys.~{\bf 19}, 575 (1978)}.
%
\bibitem{Pusz1978} 
W. Pusz and S.L. Woronowicz, \href{https://doi.org/10.1007/BF01614224}{Comm. Math. Phys.~{\bf 58}, 273 (1978)}.
%
\bibitem{Lorch18}
N. L\"orch, C. Bruder, N. Brunner, and P.P. Hofer,
\href{https://doi.org/10.1088/2058-9565/aacbf3}{Quantum Sci. Technol.~{\bf 3} 035014 (2018)}.
%
\bibitem{Plastina17}
G. Francica, J. Goold, F. Plastina, and M. Paternostro,
\href{https://doi.org/10.1038/s41534-017-0012-8}{npj Quantum
Information~{\bf 3}, 12 (2017)}.
%
\bibitem{Scarani18}
S. Seah, S. Nimmrichter, and V. Scarani,
\href{https://doi.org/10.1088/1367-2630/aab704}{New J. Phys.~{\bf 20}, 043045 (2018)}.
%
\bibitem{Kurizki18}
W. Niedenzu, V. Mukherjee, A. Ghosh, A.G. Kofman, and G. Kurizki,
\href{https://doi.org/10.1038/s41467-017-01991-6}{Nat. Comm.~{\bf 9}, 165 (2018)}.
%
\bibitem{Stockklauser17} 
A. Stockklauser, P. Scarlino, J.V. Koski, S. Gasparinetti, C.K. Andersen, C. Reichl, W. Wegscheider, T. Ihn, K. Ensslin, and A. Wallraff, 
\href{http://dx.doi.org/10.1103/PhysRevX.7.011030}{Phys. Rev. X~{\bf 7}, 011030 (2017)}.
%
\bibitem{Samkharadze18}
 N. Samkharadze, G. Zheng, N. Kalhor, D. Brousse, A. Sammak, U.C. Mendes, A. Blais, G. Scappucci, and L.M.K. Vandersypen, \href{http://dx.doi.org/10.1126/science.aar4054}{Science~{\bf 25}, eaar4054 (2018)}.
%
\bibitem{WallsMilburn2007}
D.F. Walls and G.J. Milburn, \href{https://doi.org/10.1007/978-3-540-28574-8}{\textit{Quantum Optics}} (Springer Science \& Business Media, 2007). 
%
\bibitem{SOM}
See Supplemental Material File, which includes Refs.\cite{Holdstein40,Binder15T}.
%
%
\bibitem{Holdstein40}
T. Holstein and H. Primakoff, \href{http://doi:10.1103/PhysRev.58.1098}{Phys. Rev.~{\bf 58}, 1098 (1940)}. 
%
\bibitem{Binder15T} 
F. C. Binder, S. Vinjanampathy, K. Modi, and J. Goold, 
\href{http://dx.doi.org/10.1103/PhysRevE.91.032119}{Phys. Rev. E {\bf 91}, 032119 (2015).}
%
\bibitem{goold_prb_2015}
J. Goold, C. Gogolin, S.R. Clark, J. Eisert, A. Scardicchio, and A. Silva, \href{https://doi.org/10.1103/PhysRevB.92.180202}{Phys. Rev. B~{\bf 92}, 180202(R) (2015)}.
%
\bibitem{nandkishore_annurev_2015}
R. Nandkishore and D.A. Huse, \href{https://doi.org/10.1146/annurev-conmatphys-031214-014726}{Annu. Rev. Condens. Matter Phys.~{\bf 6}, 15 (2015)}.
%
\bibitem{eisert_rmp_2010}
J. Eisert, M. Cramer, and M.B. Plenio, \href{https://doi.org/10.1103/RevModPhys.82.277}{Rev. Mod. Phys.~{\bf 82}, 277 (2010)}.
%
\bibitem{bogoliubov_jphysA_1996}
N.M. Bogoliubov, R.K. Bullough, and J. Timonen, \href{https://doi.org/10.1088/0305-4470/29/19/015}{J. Phys. A: Math. Gen.~{\bf 29}, 6305 (1996)}.
%
\bibitem{Bastarrachea11}
M.A. Bastarrachea-Magnani and J.G. Hirsch, \href{https://rmf.smf.mx/pdf/rmf-s/57/3/57_3_69.pdf}{Rev. Mex. Fis. S~{\bf 57}, 69 (2011)}.
%
\bibitem{Nielsen_and_Chuang}
M.A. Nielsen and I.L. Chuang, \href{https://doi.org/10.1017/CBO9780511976667}{{\it Quantum Computation and Quantum Information}} (Cambridge University Press, Cambridge, England, 2000). 
%
%
\bibitem{Dicke54} 
R.H. Dicke, \href{http://dx.doi.org/10.1103/PhysRev.93.99}{Phys. Rev.~{\bf 93}, 99 (1954).}
%
\bibitem{QuTip}
J.R. Johansson, P.D. Nation, and F. Nori, \href{https://doi.org/10.1016/j.cpc.2012.11.019}{Comp. Phys. Comm.~{\bf 184}, 1234 (2013)}.
%
\end{thebibliography}

\begin{thebibliography}{78}
%
\bibitem{Holdstein40S}
T. Holstein and H. Primakoff, \href{http://doi:10.1103/PhysRev.58.1098}{Phys. Rev.~{\bf 58}, 1098 (1940)}. 
%
\bibitem{Nielsen_and_ChuangS}
M.A. Nielsen and I.L. Chuang, \href{https://doi.org/10.1017/CBO9780511976667}{{\it Quantum Computation and Quantum Information}} (Cambridge University Press, Cambridge, England, 2000). 
%
\bibitem{Binder15TS} 
F. C. Binder, S. Vinjanampathy, K. Modi, and J. Goold, 
\href{http://dx.doi.org/10.1103/PhysRevE.91.032119}{Phys. Rev. E {\bf 91}, 032119 (2015).}

%
\bibitem{Dicke54S} 
R.H. Dicke, \href{http://dx.doi.org/10.1103/PhysRev.93.99}{Phys. Rev.~{\bf 93}, 99 (1954).}
%
%
\bibitem{Ferraro17S} 
D. Ferraro, M. Campisi, G.M. Andolina, V. Pellegrini, and 
M. Polini, \href{https://dx.doi.org/110.1103/PhysRevLett.120.117702} {Phys. Rev. Lett.~{\bf 120}, 117702 (2018)}.
%
\bibitem{Andolina18S} 
G.M. Andolina, D. Farina, A. Mari, V. Pellegrini, V. Giovannetti, and M. Polini,\href{https://doi.org/10.1103/PhysRevB.98.205423} {Phys. Rev. B~{\bf 98 }, 205423 (2018)}.
%

%
\end{thebibliography}
\end{document}